\documentclass[a4paper,11pt]{article}
\pdfoutput=1 % if your are submitting a pdflatex (i.e. if you have
\usepackage{jcappub} % for details on the use of the package, please
\usepackage[T1]{fontenc} % if needed
\usepackage{gensymb}

\title{Dark matter substructure cannot explain properties of the \textit{Fermi} Galactic Centre excess}

\author[a,1]{Hamish A. Clark,\note{Corresponding author.}}
\author[b]{Pat Scott,}
\author[b,c]{Roberto Trotta,}
\author[a]{and Geraint F. Lewis}

\affiliation[a]{Sydney Institute for Astronomy, School of Physics A28,\\ The University of Sydney, NSW 2006, Australia}
\affiliation[b]{Department of Physics, Imperial College London,\\Blackett Laboratory, Prince Consort Road, London SW7 2AZ, UK}
\affiliation[c]{Data Science Institute,\\William Penney Laboratory, Imperial College London, London SW7 2AZ}

\emailAdd{hamish.clark@sydney.edu.au}

\abstract{An excess of gamma rays has been identified at the centre of the Milky Way, and  annihilation of dark matter has been posited as a potential source. This hypothesis faces significant challenges: difficulty characterizing astrophysical backgrounds, the need for a non-trivial adiabatic contraction of the inner part of the Milky Way's dark matter halo, and recent observations of photon clustering, which suggest that the majority of the excess is due to unresolved point sources.  Here we point out that the apparent point-like nature of the emission rules out the dark matter interpretation of the excess entirely. Attempting to model the emission with dark matter point sources either worsens the problem with the inner slope, requires an unrealistically large minihalo fraction toward the Galactic Centre, or overproduces the observed emission at higher latitudes.}

\begin{document}
\maketitle
\flushbottom

\section{Introduction}

An excess of high-energy gamma rays has been observed toward the Galactic Centre (GC) by the \textit{Fermi} Large Area Telescope (\textit{Fermi}-LAT; \cite{FermiGC,Fermi2017}). This excess is not easily explained by known astrophysical sources. It peaks at energies of $\sim$2\,GeV, appears spherically distributed, extends up to 1.5\,kpc from the GC, and falls steeply with distance from the GC, exhibiting a profile that goes as $r^{-\Gamma}$, $\Gamma \sim 2.2 \text{--} 2.8$ \cite{Abazajian2012, Macias2014,Calore2015b, Daylan2016}. Proposed explanations include cosmic ray injection \cite{Carlson2014, Petrovic2014, Cholis2015, Gaggero2015}, a population of unresolved millisecond pulsars \citep{Abazajian2011,Mirabal2013,Abazajian2014,Bhakta2017,Ploeg2017}, or the self-annihilation of dark matter (DM) within the Galactic halo \citep{ Hooper2011a,Hooper2011b,Calore2015a}. It has also been recently suggested that there is a correlation between the distribution of the excess and the stellar population of the Galactic Bulge \cite{Macias2016, Bartels2017}.

Among these, DM self-annihilation is of particular interest, as it would allow a characterization of the particle nature of DM. Should it annihilate, DM may produce observable radiation from the direction of the GC. For particular annihilation final states, this explanation has been found to be an excellent fit to all of the spectral and morphological properties of the observed excess \citep{Calore2015a, Eiteneuer2017,Achterberg2017}.  However, recent analyses of \textit{Fermi}-LAT photon map statistics \citep{Bartels2016,Lee2016} have suggested that the vast majority of the excess originates from unresolved point sources. Although the spectrum of the apparent point-like emission has not yet been shown to match that of the observed excess, the fluxes are very similar, providing weight to the millisecond pulsar hypothesis.

The DM halo of the Milky Way is expected to contain a population of subhalos. The exact nature and abundance of the substructure is unknown. However, cold dark matter simulations predict the existence of small-scale structure \citep{Aquarius}, which should theoretically exist right down to the DM free-streaming scale. If DM annihilates, these substructures would provide a significant boost to the observed annihilation rate \citep{Kamionkowski2010,Anderhalden2013,Bartels2015}, contributing substantially to the overall gamma-ray emission observed from the GC.

In this paper we investigate the possibility that the unresolved point sources identified by \cite{Bartels2016} and \cite{Lee2016} may be small-scale DM halos. This scenario could potentially rescue the DM interpretation of the excess, by remaining compatible with observations of photon clustering that indicate a point-source origin for the emission. In order to determine the viability of such a scenario, we investigate the morphology and implied photon statistics of this boosted signal, using \textit{Fermi}-LAT observations to constrain the model parameters. For all substructure cases that we consider, we find that the signal can be explained by the presence of DM substructure only if the inner slope of the Galaxy's DM halo is drastically steepened by adiabatic contraction, or if the concentration of subhalos increases substantially toward the GC.  The parameter values that this requires are so different to results obtained from state-of-the-art numerical simulations that we conclude substructure considerations rule out a dark matter interpretation of the excess.

\section{Structure \& Substructure}
\label{section:substructure}
We model the density profile of the smooth DM halo of the Galaxy with the generalized Navarro-Frenk-White (NFW) profile \citep{Navarro1996,Wyithe2001},
\begin{equation}
\rho(r) = \frac{\rho_0}{\left(\frac{r}{r_s}\right)^\gamma\left(1+\frac{r}{r_s}\right)^{3-\gamma}},
\end{equation}
where $\rho_0$ is fixed by the local DM density at the position of the Sun ($\rho_\chi = 0.3~{\rm GeV\ cm}^{-3}$, at $r_\odot = 8~{\rm kpc}$), $r_s \approx 20 ~{\rm kpc}$ is the scale radius of the Galaxy \citep{JBH2016}, and $\gamma$ is the inner slope of the halo.

Given the difference in their typical formation histories, low-mass substructures have different characteristics to their large-scale counterparts. As such, the properties of DM substructure must be decoupled from those of the Galactic halo. In order to cover a wide range of substructure properties, we consider two bracketing cases in what follows: NFW subhalos and ultracompact minihalos (UCMHs). These exemplify the range of subhalos that could potentially exist at the Galactic Centre. Here the UCMH case is representative of compact tidally stripped halos, while our NFW case represents that of more diffuse subhalo structures.

In order to quantify the properties of the substructure in a simple manner, we make the assumption that all subhalos are of the same mass, $M_{\rm h}$. Given that the spatial distribution of the properties of low-mass subhalos are poorly constrained, this is the simplest assumption available.  In fact, as we showed in Ref.\ \citep{Clark16a}, the boost factor from more compact subhalos is completely independent of their mass; we have also checked that incorporating more complex mass distributions (a uniform or power-law mass distribution) does not significantly alter our results with shallower subhalo profiles either.

\subsection{NFW Subhalos}
N-body simulations predict small-scale substructure down to their smallest resolvable scale, with densities that appear to follow the NFW profile \citep{Aquarius,Jiang2016,Chua2016}. We model NFW substructure with this density profile, with inner slope $\alpha$, as
\begin{equation}
\rho_{\rm h}(r) = \frac{\delta_c \rho_c}{\left(\frac{r}{r_s}\right)^\alpha\left(1+\frac{r}{r_s}\right)^{3-\alpha}},
\end{equation}
where
\begin{align}
&\delta_c = \frac{200}{3}\frac{c_{200}^3}{\Phi(c_{200})},\label{eq:conc}\\
&\Phi(x) = \frac{x^{3-\alpha}}{3-\alpha} ~{}_2F_1(3-\alpha,3-\alpha;4-\alpha;-x),
\end{align}
and $_2F_1(a,b;c;d)$ is the Gaussian hypergeometric function, $\rho_c = 3H_0^2/8\pi G$ is the critical density of the Universe today, $H_0$ is the present-day value of the Hubble constant, $G$ is the gravitational constant, and $c_{200}$ is the halo concentration parameter, taken as a function of subhalo mass following \cite{Correa2015}. The scale radius may be calculated as a function of the halo mass by
\begin{equation}
r_s = \left(\frac{3M_h}{800\pi \rho_c c_{200}^3}\right)^{1/3}.
\end{equation}

\subsection{Ultracompact Minihalos}
Large amplitude density fluctuations in the early Universe ($10^{-3} \lesssim \delta \lesssim 0.3$) lead to an increased production of dense small-scale structures, known as ultracompact minihalos (UCMHs; \cite{Ricotti2009, Scott2009, Berezinsky2012,Berezinsky2013}). Should DM annihilate, these would be strong sources of annihilation products \cite{Scott2009}, and therefore strong probes of small-scale cosmology \cite{JG10,Bringmann2012,Shandera12,Aslanyan16}. In what follows, we consider UCMH substructure as summarized in \cite{Bringmann2012}. The density profile of a UCMH at redshift $z=0$ may be calculated as
\begin{equation}
\rho_{\rm h}(r) = \frac{3f_\chi M_{\rm h}}{16\pi R_{\rm h}^{3/4} r^{9/4}},
\end{equation}
where $f_\chi$ is the fraction of matter that is cold dark matter, $M_{\rm h}$ is the mass of the halo, and $R_{\rm h}$ is the radius of the halo,
\begin{equation}
\frac{R_{\rm h}}{\rm pc} = 1.73\left(\frac{M_{\rm h}}{M_\odot}\right)^{1/3}.
\end{equation}

UCMH density profiles are expected to soften in the innermost regions, at radii smaller than
\begin{align}
r_{\rm c} = \max \left(r_{\rm min}, r_{\rm cut} \right).
\end{align}
Here $r_{\rm c}$ is the greater of the effective annihilation radius
\begin{equation}
r_{\rm cut} = (\kappa\Delta t\langle\sigma v\rangle/m_\chi)^{4/9},
\end{equation}
given as the extent of the inner region that is annihilated away over the UCMH lifetime $\Delta t$ (which we take to be the time since equality), and the angular momentum radius $r_{\rm min}$,
\begin{equation}
r_{\rm min} \approx 2.9 \times 10^{-7} R_{\rm h} \left(\frac{1000}{z_c+1}\right)^{2.43}\left(\frac{M_{\rm h}}{M_\odot}\right)^{-0.06},
\end{equation}
which is the inner radius at which the radial infall approximation becomes inappropriate.

\section{Flux from Dark Matter Self-Annihilation}
\label{section:DMA}
The presence of substructure within the Galaxy substantially increases the rate at which DM annihilates. The addition of even a small population of overdensities can boost annihilation by orders of magnitude, depending upon the nature and prevalence of the subhalos. Here we calculate the spatially-averaged total flux produced by a substructure population embedded within the smooth DM halo of the Galaxy.

For a spherically symmetric halo at a distance $d > R_{\rm h}$ the total gamma-ray flux due to DM annihilation, differential in energy, is given by
\begin{equation}
\mathcal{F}_h(d,E) = \sum_k \frac{\textrm{d}N_k}{\textrm{d}E} \frac{\langle \sigma_k v\rangle}{2d^2m_\chi^2} \int_0^{\rm R_h} \rho_h^2(r) r^2 \textrm{d}r,
\end{equation}
where $\rho_{\rm h}(r)$ is the DM density at a distance $r$ from the centre of the halo, $m_\chi$ is the DM particle mass, $R_{\rm h}$ is the maximum radius of the halo, and $dN_k/dE$ and $\langle \sigma_k v\rangle$ are the differential photon yield and cross section from the $k$th annihilation channel, respectively.

\begin{figure*}
\includegraphics[width=0.9\textwidth]{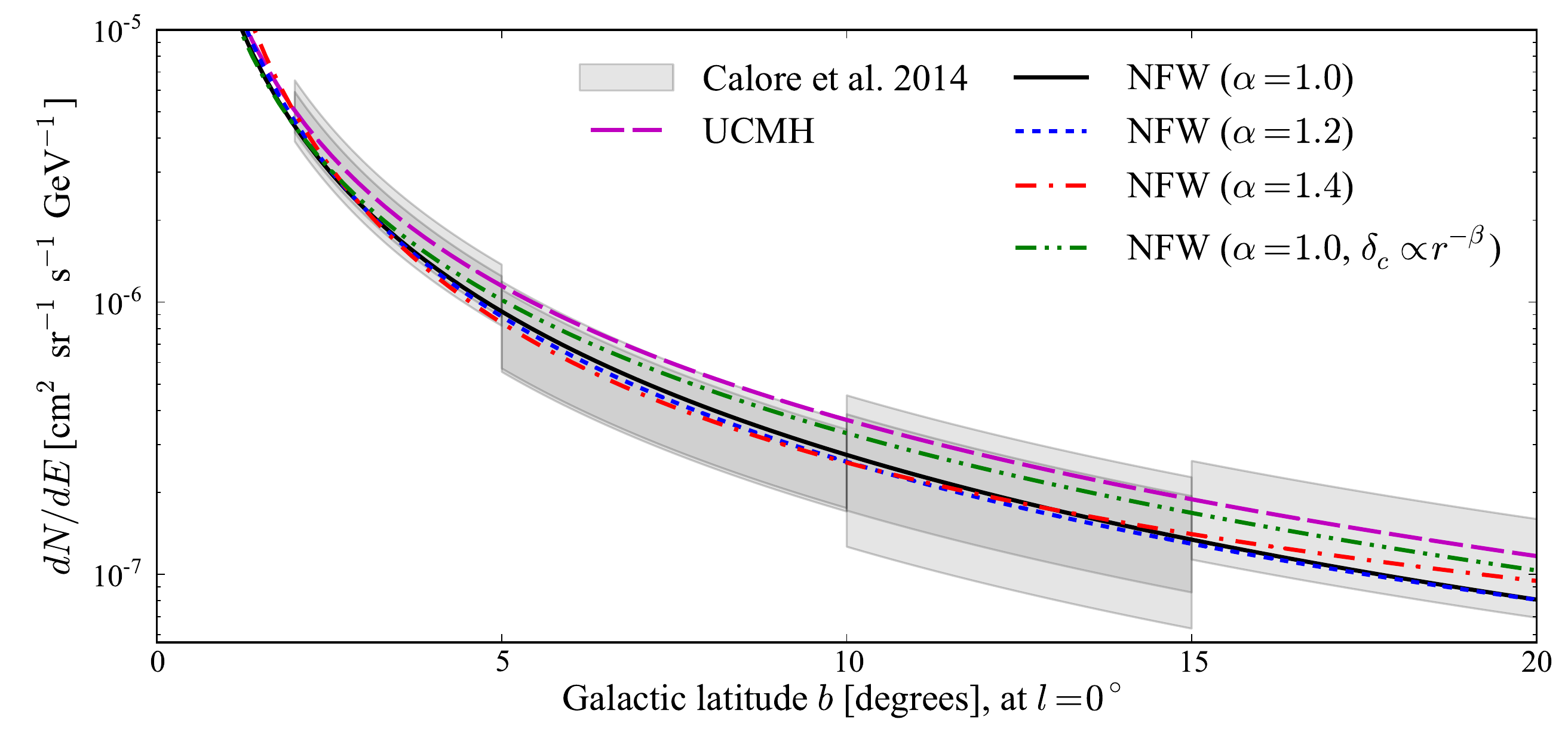}
\caption{Radial differential flux profile of the \textit{Fermi} gamma ray excess at 2 GeV, with posterior mean DM annihilation flux profiles for a range of different substructure models. The fluxes within each region of interest (ROI) from Ref.\ \cite{Calore2015b} are shown as 68\% confidence level bands, plotted as $\mathcal{F} \propto r^{-2.7}$ within each ROI to guide the eye, and including statistical and systematic errors. Overlapping regions correspond to the north and south regions of the sky. Note that for clarity, we do not plot the two ROIs at $l\ne0\degree$ included in our fit (ROIs 7 \& 8). The ROI definitions can be found in Table \protect\ref{Table2}, and the parameters of the plotted models are given in Table \protect\ref{Table1}.}
\label{fig:fluxplot}
\end{figure*}

\begin{table}[]
\centering
\begin{tabular}{lllc}
ROI & Radial cut & Additional cut & $\Omega_{\rm ROI}$ (sr) \\\hline
1   &  $\sqrt{l^2 + b^2} < 5\degree$ & $2\degree < b$  & $6.0 \times 10^{-3}$ \\
2   &  $\sqrt{l^2 + b^2} < 5\degree$ & $2\degree < -b$  & $6.0 \times 10^{-3}$ \\
3   &  $5\degree < \sqrt{l^2 + b^2} < 10\degree$ & $|l| < b$  & $1.8 \times 10^{-2}$ \\
4   &  $5\degree < \sqrt{l^2 + b^2} < 10\degree$ & $|l| < -b$  & $1.8 \times 10^{-2}$ \\
5   &  $10\degree < \sqrt{l^2 + b^2} < 15\degree$ & $|l| < b$  & $2.9 \times 10^{-2}$ \\
6   &  $10\degree < \sqrt{l^2 + b^2} < 15\degree$ & $|l| < -b$  & $2.9 \times 10^{-2}$ \\
7   &  $5\degree < \sqrt{l^2 + b^2} < 15\degree$ & $2\degree < |b| < l$  & $3.5 \times 10^{-2}$ \\
8   &  $5\degree < \sqrt{l^2 + b^2} < 15\degree$ & $2\degree < |b| < -l$  & $3.5 \times 10^{-2}$ \\
9   &  $15\degree < \sqrt{l^2 + b^2} < 20\degree$ & $2\degree < |b|$  & $1.5 \times 10^{-1}$ \\
\end{tabular}
\caption{Parameters in Galactic co-ordinates of the regions of interest (ROIs) that we use in our analysis. From Ref.\ \cite{Calore2015b}.}
\label{Table2}
\end{table}

\begin{table}[]
\centering
\begin{tabular}{lrrrrrr}
Scan                              & $\gamma$             & $M_{\rm h}\, (M_\odot)$ & $\log_{10} \left[\frac{\langle \sigma v\rangle}{m_{\chi}^2}\cdot \frac{\mathrm{GeV}^2\,\mathrm{s}}{\mathrm{cm}^3}\right]$ & $f$                  & $\log_{10}A$         & $\beta$              \\ \hline
UCMH                              & 2.61                 & 0.50                & $-$36.97                                         & 0.558                & ---                  & ---                  \\
NFW, $\alpha = 1.0$               & 1.32                 & 0.50                & $-$29.34                                         & 0.934                & ---                  & ---                  \\
NFW, $\alpha = 1.2$               & 1.38                 & 0.28                & $-$29.49                                         & 0.931                & ---                  & ---                  \\
NFW, $\alpha = 1.4$               & 1.53                 & 0.47                & $-$29.83                                         & 0.922                & ---                  & ---                  \\
NFW, $\delta_c\propto r^{-\beta}$ & 2.06                 & 0.07                & $-$32.74                                         & 0.666                & 8.570                & 0.524               \\
                                  & \multicolumn{1}{l}{} & \multicolumn{1}{l}{}  & \multicolumn{1}{l}{}                           & \multicolumn{1}{l}{} & \multicolumn{1}{l}{} & \multicolumn{1}{l}{}
\end{tabular}
\caption{Parameter vales for each of the posterior mean solutions with fluxes plotted in Fig.\ \ref{fig:fluxplot}.}
\label{Table1}
\end{table}

Likewise, the flux per unit volume from the smooth DM component with local density $\rho_\chi$ at a distance $d$ may be found as
\begin{equation}
\mathcal{F}_{\rm smooth}(d,E) = \sum_k \frac{dN}{dE} \frac{\langle \sigma_k v\rangle}{2d^2m_\chi^2} \left[(1-f)\rho_\chi\right]^2,
\end{equation}
where $f$ is the fraction of DM that is contained within substructure.

Additionally, there is a contribution from the annihilation of DM particles within the smooth component with those within the overdense region of the subhalo:
\begin{equation}
\mathcal{F}_{\rm cross}(d,E) = \sum_k \frac{\textrm{d}N_k}{\textrm{d}E} \frac{\langle \sigma_k v\rangle}{2d^2m_\chi^2} \int_0^{\rm R_h} 2(1-f)\rho_\chi \rho_h(r) r^2\textrm{d}r.
\end{equation}

The total gamma-ray flux per steradian from any point in the sky may then be found as
\begin{align}
\frac{\textrm{d}\mathcal{F}}{\textrm{d}\Omega}(E) = \int_0^{d_{\rm max}} d'^2\big[\mathcal{F}_{\rm smooth}&(d',E) + \mathcal{F}_{\rm cross}(d',E)n(d') \nonumber\\&+ \mathcal{F}_h(d',E) n(d')\big] dd',
\label{totalflux}
\end{align}
where $n = f\rho_\chi/f_\chi M_{\rm h}$ is the local number density of halos at distance $d'$ along the line of sight, and $d_{\rm max}$ is evaluated at the virial radius of the Galaxy's DM halo, taken as $R_{\rm vir} = 360~{\rm kpc}$.

By computing the expected flux from subhalo-subhalo collisions as a function of impact parameter, and performing a nearest-neighbour analysis, we checked that the contribution of cross-annihilation between colliding subhalos is subdominant. Our tests indicate that, depending on $f$, including this term could increase or decrease the slope of the flux profile, by providing a relative boost at higher or lower latitudes of up to 25\%; this is within the systematic error band of the observations that we use for constraining substructure models.

\section{Statistical Analysis}
To determine the substructure properties required to explain the \textit{Fermi} excess, we express the gamma-ray flux in terms of the parameters of the annihilation and subhalo structure models described in the previous sections. We constrain these quantities using summary statistics of both the morphology of the gamma-ray excess and the photon arrival directions. We assume 100\% annihilation of an $m_\chi = 100$\,GeV DM particle into $b\bar{b}$ final states, and allow the overall annihilation cross-section $\langle \sigma v \rangle$ to float. This model does give a reasonable fit to the observed spectrum of the excess, but because we do not perform a spectral fit to the gamma-ray data, the assumed DM model has minimal impact on our results.

To constrain the morphological properties of the excess, we use the results of the analysis of the GC gamma-ray signal by \cite{Calore2015b}. We fit to the observed excess flux (differential in energy, measured at 2\,GeV) integrated within each of the 9 innermost regions of interest (ROI), neglecting the outermost region due to overlap with the \textit{Fermi} bubbles. The exact definitions of these ROIs can be found in Table \ref{Table2}.
We approximate the reported values and errors for these regions as Gaussian. These 9 regions encapsulate the morphology of the excess flux both above and below the Galactic Plane. We compare these to the modelled flux by integrating Eq.\ \eqref{totalflux} over each ROI for a given substructure model and set of model parameters ($f$, $M_{\rm h}$, $\langle \sigma v\rangle$ and $\gamma$).  Note that $\gamma$ here is the slope parameter of the Galactic halo, not of the minihalos.  We leave $\gamma$ as a free parameter in each of our fits, assuming a fixed minihalo slope, $\alpha$ --- carrying out multiple fits with different values of $\alpha$. We approximate the morphology likelihood as a Gaussian, given by
\begin{equation}
\mathcal{L}_{\rm morph} = \prod_{i=1}^{\rm 9} \exp\left[-\frac{(\mathcal{F}_i-\mu_i)^2}{2\sigma_i^2}\right],
\end{equation}
where $i$ is the ROI index, $\mu_i$ is the mean flux at 2\,GeV estimated by \cite{Calore2015b}, $\mathcal{F}_i$ is the corresponding prediction from Eq.\ \eqref{totalflux}, and $\sigma_i$ is the sum in quadrature of the (correlated) systematic and (uncorrelated) statistical error. Note that we neglect the correlations between systematic errors in different ROIs; this is a conservative choice, as it reduces our ability to rule out substructure models.

Statistical studies of the \textit{Fermi} photon map have suggested that $\gtrsim 95\%$ of the excess originates from a population of unresolved point sources. Here we assume that any unresolved point-like emission comes from small-scale DM substructures. To include this in our analysis, we use the results from \cite{Lee2016}. Their `non-Poissonian template fit' differentiates between Poissonian and non-Poissonian photon statistics, to distinguish diffuse and point-like emission. We take the posterior distributions for the total flux from the point-like ($I^{\rm NFW}_{\rm PS}$) and smooth components ($I_{\rm NFW}$) from the region within $10\degree$ of the GC, with $|b|\geq2\degree$. To avoid using the data twice (i.e., the total flux is already included in the morphology likelihood) we use the posterior for their ratio (with 3FGL sources masked), $z \equiv x/y \equiv I_{\rm NFW} / I^{\rm NFW}_{\rm PS}$. The posterior for the ratio is given by \cite{Curtiss1941}
\begin{equation}
P(z|D) = \int_{-\infty}^\infty |y| P(zy,y|D) dy,
\label{ratio}
\end{equation}
where $P(x,y|D)$ is the 2D joint posterior of the smooth and point-like fluxes, given {\em Fermi} data $D$. Given the posteriors for $I^{\rm NFW}$ and $I^{\rm NFW}_{\rm PS}$ are uncorrelated in the results of \cite{Lee2016}, we construct $P(x,y|D)$ from the individual 1D distributions in Fig.\ S4 of that paper. We reinterpret this posterior as a likelihood function and denote it $\mathcal{L}_{\rm PS}(z)$, and we use $z = \mathcal{F}_{\rm smooth}/(\mathcal{F}_{\rm h}+\mathcal{F}_{\rm cross})$, where $\mathcal{F}$ is the total flux predicted within the Lee et al. ROI (\cite{Lee2016}; $r\le10\degree$, $|b|\geq2\degree$) according to Eq.\ \eqref{totalflux}.

We hasten to point out that our treatment of the fraction of the excess flux that must be attributed to point-like structures is by its nature somewhat approximate. The finding of Ref.\ \cite{Lee2016} that more than $\sim$95\% of the excess can be attributed to unresolved point sources relies on an assumed form of the point source luminosity function, which might be expected to be somewhat more strongly weighted towards lower luminosities for DM subhalos than for the population of pulsars that the authors of Ref.\ \cite{Lee2016} had in mind.  In this sense, directly adopting the results of Ref.\ \cite{Lee2016} in our analysis is conservative, as the equivalent lower bound on the fraction of the excess emission attributable to \textit{dark matter} point sources, were the authors of Ref.\ \cite{Lee2016} to repeat their analysis with such a substructure model, would be even larger than 95\%.  Similarly, the analysis of \cite{Bartels2016} only finds \textit{direct} evidence for a small fraction of the excess flux being attributable to point sources below the standard \textit{Fermi} significance threshold, and also requires some form of extrapolation to lower point-source luminosities in order to determine the \textit{total} fraction of the flux attributable to point-like and smooth emission.  Significant uncertainty therefore remains on the exact fraction of the excess flux that must come from point-like emitters.  Nonetheless, given the size of the systematic uncertainties associated with the extraction of the excess itself, these are acceptable uncertainties for our purposes.

The joint posterior for the parameters of our model, $\Theta$, is given by (up to an irrelevant normalization constant)
\begin{equation}
P(\Theta | d) \propto \mathcal{L}_{\rm PS} \mathcal{L}_{\rm morph}P(\Theta)
\end{equation}
where $P(\Theta)=P(\gamma)P(M_h)P(\langle \sigma v \rangle)P(f)$ is the prior distribution for the parameters. We adopt uniform priors on $\gamma \in [0,3]$, $f \in [0,1]$, $\log_{10} (M_{\rm h}/M_\odot) \in [-12,9]$ and $\log_{10}(\langle \sigma v\rangle/{\rm cm}^3 ~{\rm s}^{-1}) \in [-40,-20]$. We use a nested sampling procedure to sample from the posterior distribution and infer the parameters of substructure models.

\section{Substructure as the source of the excess}

Assuming substructure consists of NFW subhalos, we performed scans for three different choices of the subhalo inner slope, $\alpha$. We show the flux profiles for the resulting posterior mean values of the parameters in Fig.\ \ref{fig:fluxplot}.  For these scans, we assumed that the subhalo fraction and concentration are independent of galactocentric radius; we explore the impact of allowing for a non-flat concentration-radius relation later in this Section, and a non-flat fraction-radius relation in Appendix \ref{app}.

\begin{figure}
\centering
\includegraphics[width=0.8\textwidth]{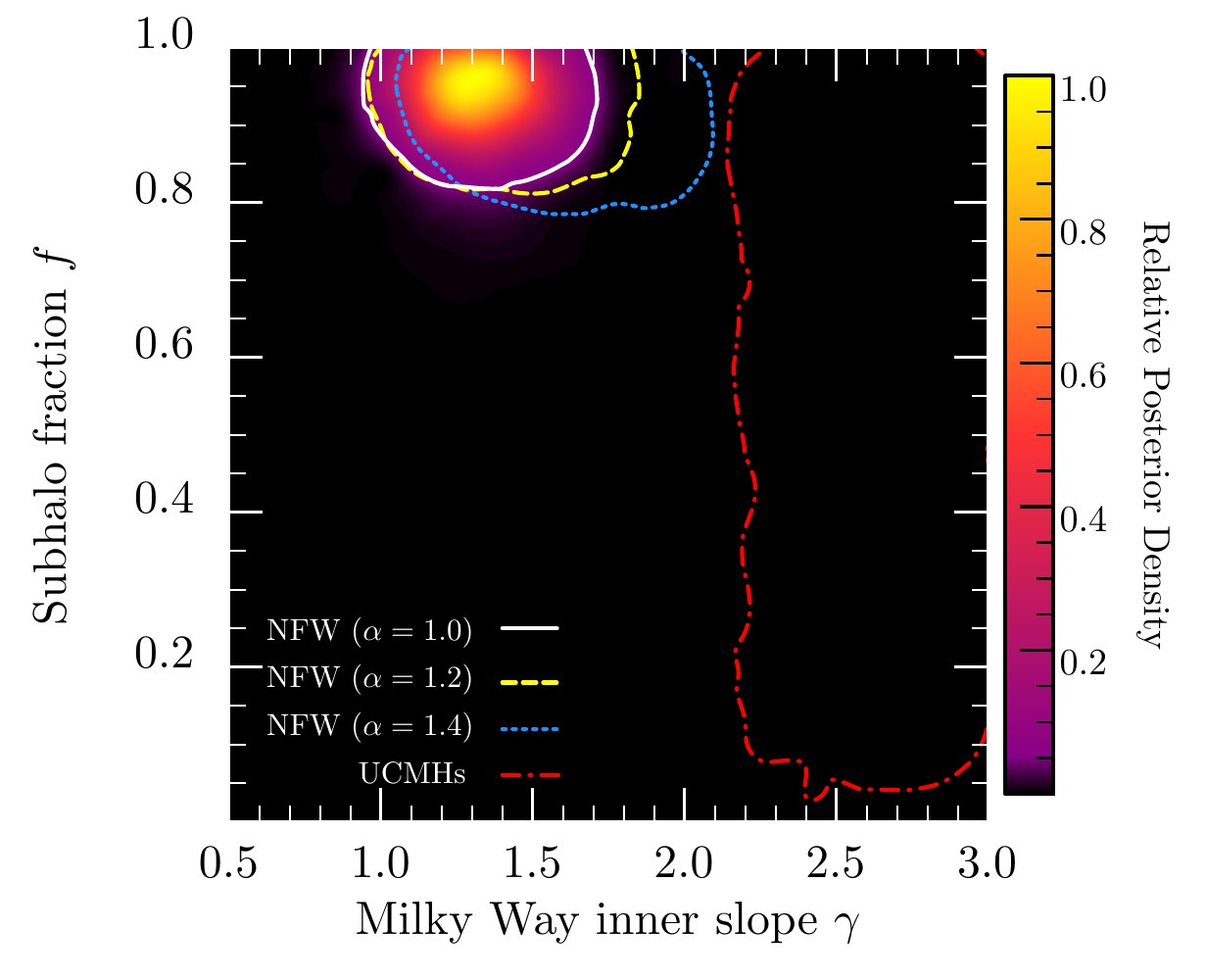}
\caption{95\% CRs for the substructure fraction and the inner slope of the Galactic DM density profile, for four different substructure models. Shading corresponds to the posterior density for the non-contracted NFW profile ($\alpha = 1.0$).}
\label{fig:f_v_gamma}
\end{figure}

While subhalo models provide reasonable morphological fits (reduced $\chi^2\in [0.24,1.0]$), this requires a substantial abundance of substructure in the GC if the point source population of \cite{Lee2016} is to be attributed to DM halos.  In Fig.\ \ref{fig:f_v_gamma}, we show 95\% (highest posterior density) credible regions (CRs) for $\gamma$ and $f$, marginalizing over $M_h$ and $\langle \sigma v\rangle$. For all three choices of $\alpha$, we find $f \gtrsim 0.8$. Given the relatively small scale of the GC, any substructure present within this region would be expected to undergo mergers and tidal disruptions. According to both analytical and numerical studies, these interactions result in a decreased amount of substructure at small Galactic radii -- estimates of the substructure fraction within 3\,kpc of the GC vary, but all studies with sufficient resolution predict $f \lesssim 0.05$ \citep{Aquarius,Diemand2007,Diemand2008e,Diemand2008N,Zhu2016,Jiang2016,Stref2016}. Given the difference between this value of $f$ and the much larger value required to explain the \textit{Fermi} excess, we conclude that the GC point source population is not constituted in any significant way by NFW subhalos.

\begin{figure}
\centering
\includegraphics[width=0.8\textwidth]{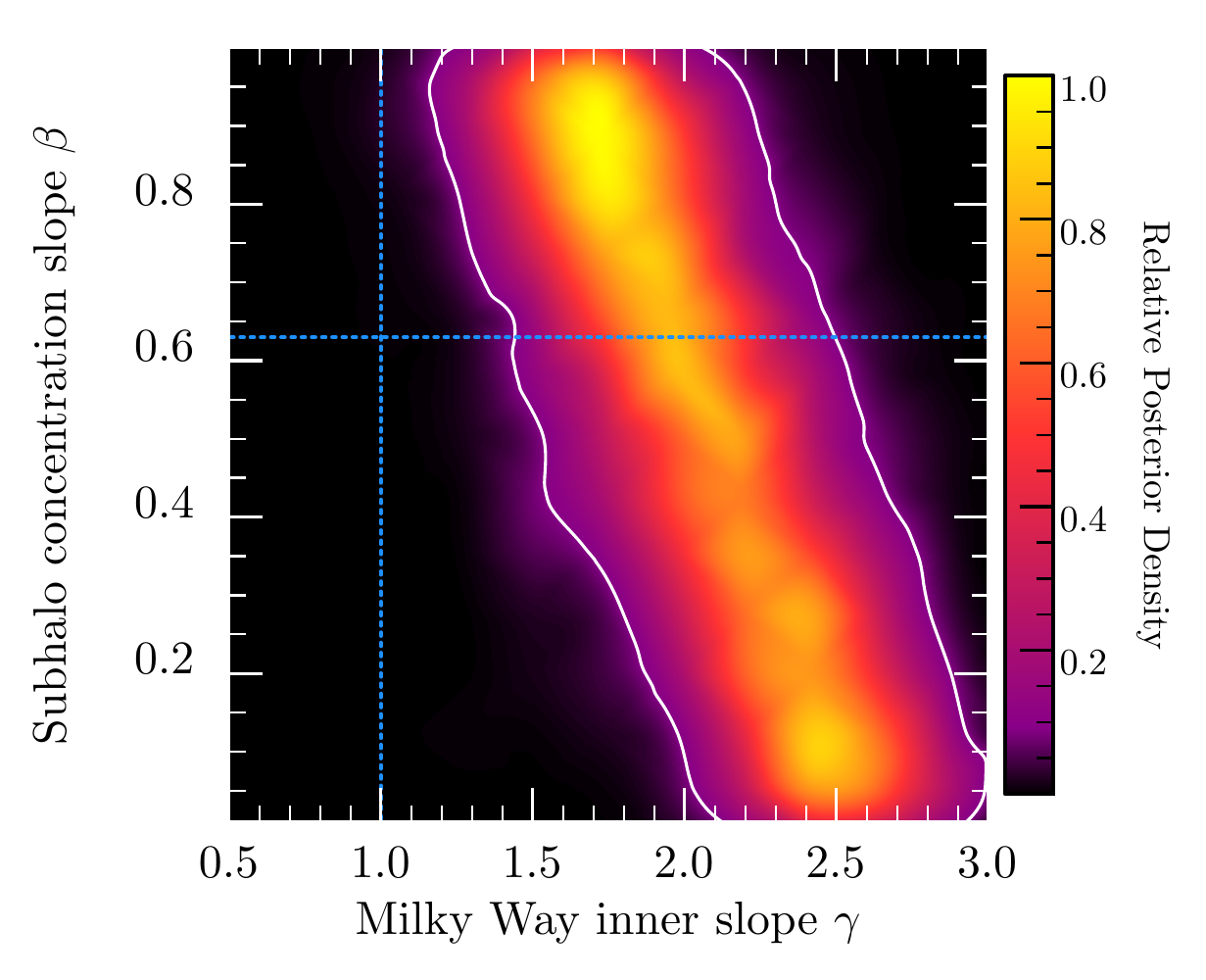}
\caption{Posterior probability density and 95\% CR for the radial slope of the subhalo concentration function and the inner slope of the Galaxy's DM density profile. Vertical and horizontal dashed lines correspond to the results of \citep{Aquarius} ($\gamma = 1$, $\beta=0.63$).}
\label{fig:beta_v_gamma}
\end{figure}

UCMHs can be strong sources of DM annihilation, meaning that a smaller number of UCMHs would produce the same signal as a greater number of NFW subhalos. We show the resulting 95\% confidence CRs on $\gamma$ and $f$ for UCMH substructure in Fig.\ \ref{fig:f_v_gamma}, marginalized over $M_h$ and $\langle \sigma v\rangle$. Although it is possible to fit the data with a far lower substructure fraction, a large increase in the inner slope of the Galactic halo is needed to explain the excess. This may be understood by considering that the annihilation rate per unit volume within a smooth halo ($f=0$) will be $\mathcal{A} \propto \rho_\chi^2 \propto r^{2\gamma}$, while that in a clumpy halo ($f=1$) goes as the number density of subhalos, $\mathcal{A}\propto \Gamma \rho_\chi \propto r^\gamma$, where $\Gamma$ is the flux emitted per subhalo. This means that if the DM halo consists predominantly of substructure, the gamma-ray excess would be much flatter toward the GC than is observed, unless the Galactic halo is significantly adiabatically contracted.\footnote{This is not an effect of favouring any particular value of $M_h$, as the UCMH substructure boost factor is in fact independent of their mass function \citep{Clark16a}. Indeed, in all scans performed here (NFW and UCMH alike) we find equal preference for all values of $\log_{10} (M_{\rm h}/M_\odot) \in [-12,9]$.}

The degree of adiabatic contraction ($\gamma$) and the substructre fraction ($f$) required to simultaneously fit the morphological and point-source likelihoods are strongly correlated -- not just with each other, but also with the central minihalo density slope ($\alpha$).  UCMHs are the most extreme case: with such large values of $\alpha$, the majority of the overall flux comes from subhalos even when $f$ is small (satisfying the point-source likelihood), and the flux due to point sources is dominated by $\mathcal{F}_\mathrm{h}$ rather than $\mathcal{F}_\mathrm{cross}$.  This leads to a preferred Milky Way slope of $\gamma \sim 2.5$, approximately twice that required to fit the morphology of the excess purely with annihilation in a smooth halo.  With increasingly shallow subhalos (smaller $\alpha$), larger subhalo fractions (larger $f$) are required for subhalos to dominate the overall emission and thereby satisfy the point-source likelihood.  At the same time, with smaller $\alpha$, an increasing fraction of the point-like emission comes from $\mathcal{F}_\mathrm{cross}$, which goes as the square of the DM density, much like the emission from the smooth halo itself, leading to steadily smaller preferred Milky Way inner slope.

So far, we have assumed that subhalos have the same properties throughout the Galaxy. An increase in subhalo abundance toward the GC would allow for a steep excess without requiring substantial adiabatic contraction of the main halo. This seems unlikely, however, as tidal interactions have been found to destroy subhalos in dense regions. These same interactions also strip the exterior regions of subhalos, resulting in an increased subhalo concentration parameter ($c_{200}$) closer to the GC. This increased concentration could in the same way allow NFW subhalos to provide a steeper excess, lowering the required value of $f$.

We investigated the effect of a spatially-varying subhalo concentration, modelled as a radial power law
\begin{equation}
\delta_c = A\left(\frac{r}{r_\odot}\right)^{-\beta},
\label{powerlaw}
\end{equation}
with $r_\odot$ the galactocentric distance of the Sun and $c_{200}$ determined from Eq.\ \eqref{eq:conc}.  We show the results of this scan in Fig.\ \ref{fig:beta_v_gamma}.  Adopting a subhalo inner slope of $\alpha=1.0$, a uniform prior on $\beta \in [0,1]$ and $\log_{10}A \in [5, 10]$, we find that the case of a regular NFW Milky Way halo with zero adiabatic contraction ($\gamma = 1.0$) is excluded with $> 99\%$ probability for $\beta<1$.  Note that by introducing the free parameter $A$, these fits not only allow for the concentration to vary with radius, but for the overall normalisation of the concentration of all minihalos to vary; in practice, this makes allowing $A$ to vary essentially equivalent to scanning over $\alpha$.

Within the Aquarius N-body simulation, it is seen that subhalos follow Eq.\ \eqref{powerlaw} with $\beta \sim 0.63$ down to a resolution of $\sim 10$\,kpc.  At this value of $\beta$, even contracted NFW profiles up to $\gamma=1.4$ are excluded with $> 95\%$ probability.  Although $\gamma > 1.4$ is observationally permitted by dynamical measurements of the Galactic Bulge \cite{Iocco17}, these data disfavour values above $\gamma=1.2$.  N-body simulations also indicate that such steep inner profiles are rather implausible.  It therefore appears that even with a radially-varying subhalo concentration, DM annihilation in subhalos is not compatible with the observed properties of the \textit{Fermi} excess.

\section{Summary \& Conclusions}

Recent results suggest that the gamma-ray excess at the Galactic Centre is produced by a population of unresolved point sources. We have investigated the idea that these point sources are small-scale DM substructures, using morphological data of the \textit{Fermi-LAT} excess, as well as the statistics of individual photon arrivals.

We found that while the morphological properties of the excess can be explained by a substructure population, a significant amount of substructure is required. For a population of uncontracted NFW subhalos, we found that $\gtrsim 80\%$ of the Galactic halo must exist as substructure, in stark contradiction with expectations from numerical simulations.

This implausibly large substructure fraction can be circumvented if the subhalos are extremely dense, as with UCMHs. These are very strong producers of annihilation products, allowing a substructure fraction as low as $f \sim 0.05$. However, this requires a substantially contracted Galactic inner slope of $\gamma \gtrsim 2.2$. Even if we allow the concentration of substructure to vary with distance from the GC, it is not possible to fit the properties of the observed excess with substructure unless $\gamma \gtrsim 1.4$.  Given that such extreme contraction is not borne out in numerical simulations, we conclude that the point sources detected via the wavelet analysis of \cite{Bartels2016} and the non-Poissonian template fit of \cite{Lee2016} are of astrophysical origin.

\acknowledgments

This paper has made use of \textsf{MultiNest} \citep{Multinest2,Multinest1,Multinest3} and \textsf{pippi} \citep{pippi}. HAC acknowledges the Australian Postgraduate Awards (APA).  PS and RT are supported by STFC (ST/K00414X/1 and ST/N000838/1), and RT by an EPSRC Pathways to Impact grant. We acknowledge the University of Sydney HPC service and the Sydney Informatics Hub for providing computational resources that have contributed to the research results reported in this paper.

\appendix

\section{Appendix}
\label{app}
\begin{figure}
\centering
\includegraphics[width=0.8\textwidth]{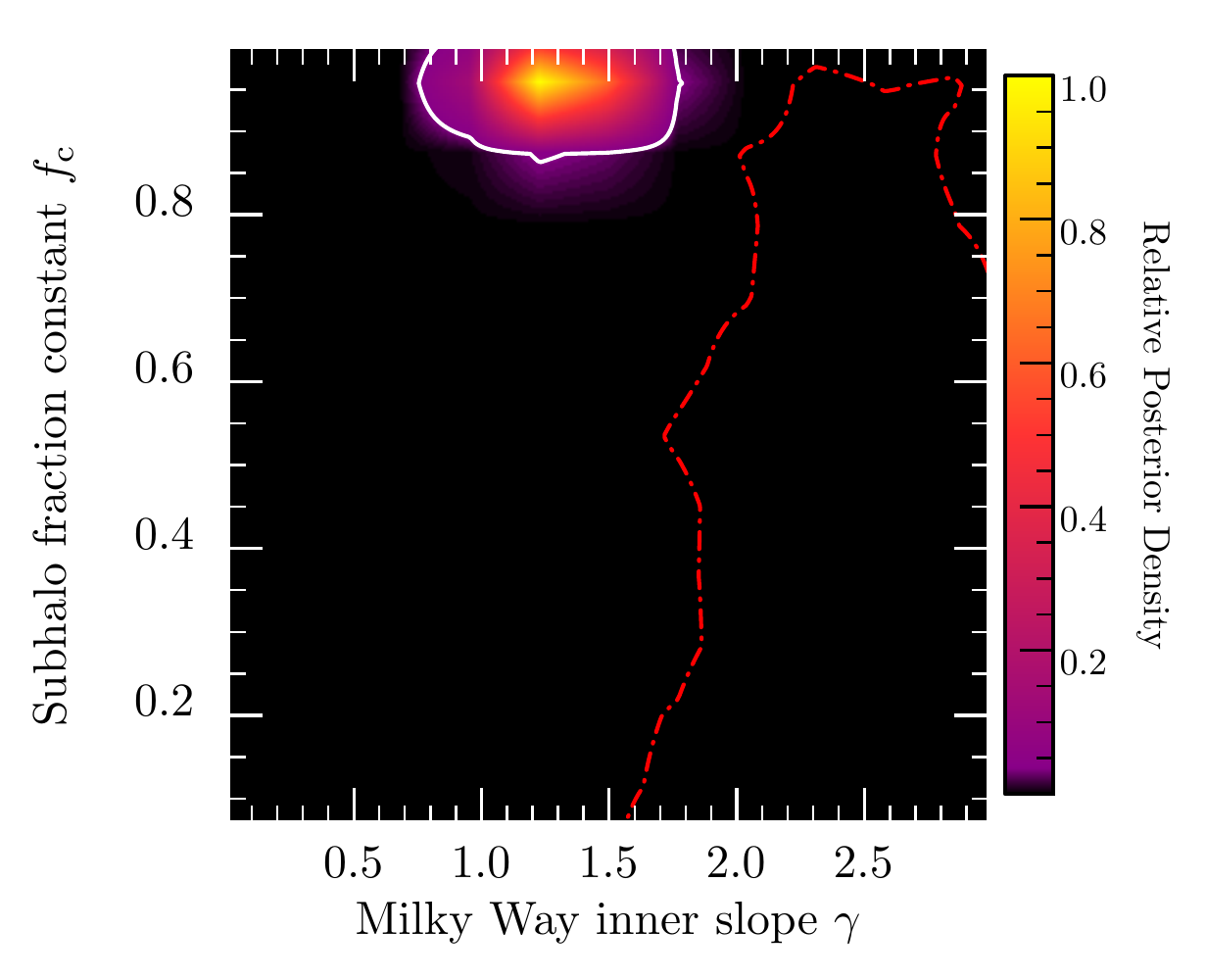}
\caption{Posterior probability densities and 95\% CRs for the subhalo fraction at $r = r_s$ and the inner slope of the Galaxy's DM density profile. The white contour is the 95\% CR for uncontracted NFW subhalos ($\alpha = 1.0$), and the red contour gives the corresponding 95\% CR for UCMH subhalos.  Shading corresponds to the posterior density for the non-contracted NFW profile.}
\label{fig:variable_f}
\end{figure}

As an additional check, we have allowed the subhalo fraction, $f$, to vary spatially, modelling it as $f=f_{\rm c}(r/r_\odot)^{\beta}$. Using a uniform prior on $\beta \in [-0.5, 0.5]$, $f_{\rm c} \in [0,1]$, we performed scans over our likelihood function.  We give the resulting posterior probability densities for both UCMHs and NFW subhalos in Fig.\ \ref{fig:variable_f}. This altered subhalo density function provided essentially the same results as we found in Fig.\ \ref{fig:f_v_gamma} for constant $f$.  This reiterates our point that substructure cannot simultaneously explain the point-like nature and morphology of the \textit{Fermi} excess.

\bibliographystyle{JHEP}
\bibliography{GCA}
\end{document}